\documentclass[5p,number]{elsarticle}
\usepackage{amsmath}
\usepackage{amsfonts}
\usepackage{lineno}

\newcommand{\degree}{\ensuremath{^\circ}}
\journal{Nuclear Instruments and Methods in Physics Research Section A}
\begin{document}
%\linenumbers
\begin{frontmatter}

\title{Search for neutrino emission in gamma-ray flaring blazars with~the~ANTARES~telescope}

%\author{Agust\'in S\'anchez Losa and Damien Dornic on behalf of the ANTARES Collaboration}
\author{Agust\'in S\'anchez Losa on behalf of the ANTARES Collaboration}
\address{agustin.sanchez@ific.uv.es, IFIC, Apartado de Correos 22085, E-46071 Valencia, Spain}

%Comments for the reviewers:
%First of all thanks a lot for the patience and for the corrections, it really need it. I'm sorry for my poor grammar and amount of mistakes, is also my first proceeding or anything close.
%Concerning to the comments of the last reviewing (Tue 06 Mar 2012):
%Q: Does this take into account times of high luminosity? In the given analysis, only ~60% of the data was usable.
%A: When I talk about a duty cycle close to 100% I'm just referring to the capacity of taking data of ANTARES. The selection of runs for the analysis, excluding the ones in less favourable conditions, but still with data taken and muon tracks reconstructed, reduce the 117 days of the studied period to an effective time of 60.8 days (the 52% of the time).
%Q: Which n_{\rm sig} is used to estimated \lambda_{\rm data} ?
%A: When the likelihood is computed, n_{\rm sig} is a free parameter to be fitted, i.e. the n_{\rm sig} is not fixed for the likelihood maximization, so is freely varied in order to maximize the likelihood estimator \lambda_{\rm data}. In average the fitted n_{\rm sig} should be around the simulated signal in the Monte Carlo simulated experiments.

\begin{abstract}
The ANTARES telescope observes a full hemisphere of the sky all the time with a duty cycle close to 100\%. This makes it well suited for an extensive observation of neutrinos produced in astrophysical transient sources. In the surrounding medium of blazars, i.\,e.\ active galactic nuclei with their jets pointing almost directly towards the observer, neutrinos may be produced together with gamma-rays by hadronic interactions, so a strong correlation between neutrinos and gamma-rays emissions is expected. The time variability information of the studied source can be obtained by the gamma-ray light curves measured by the LAT instrument on-board the Fermi satellite. If the expected neutrino flux observation is reduced to a narrow window around the assumed neutrino production period, the point-source sensitivity can be drastically improved. The ANTARES data collected in 2008 has been analysed looking for neutrinos detected in the high state period of ten bright and variable Fermi sources assuming that the neutrino emission follows the gamma-ray light curves. First results show a sensitivity improvement by a factor 2-3 with respect to a standard time-integrated point source search. The analysis has been done with an unbinned method based on the minimization of a likelihood ratio applied to data corresponding to a live time of 60 days. The width of the flaring periods ranges from 1 to 20 days. Despite the fact that the most significant studied source is compatible with background fluctuations, recently detected flares promise interesting future analyse.
\end{abstract}


\end{frontmatter}

\section{Introduction}
\label{intro}
Neutrino astronomy is an incipient field of observation of the universe. Since neutrinos do not interact electromagnetically, but weakly only, they are not absorbed and point directly to their original sources, playing the role of unique messengers. However, due to their weak interaction, neutrino telescopes are very low event rate experiments requiring high fluxes in order to detect a clear signal of a neutrino astrophysical source. Up to now no claim of an astronomy source of neutrinos further than the Sun has been made, with the exception of the well known SN1987A. Assuming a direct correlation between the neutrino emission and the light emission of a source in an hadronic acceleration scenario \cite{hadronic}, the standard analysis can be improved by a factor 2--3 in the sensitivity if we limit this search to  the periods of maximum probability of neutrino emission by the source. The time variability visible in the gamma-ray light curves measured by the Fermi-LAT \cite{fermi} can be used for this purpose.

\subsection{The ANTARES telescope}
\label{detector}
The ANTARES detector is a neutrino telescope placed at the bottom of the Mediterranean Sea (42\degree48~N, 6\degree10~E), at a depth of 2475~m, connected by a submarine cable of 42~km to the shore in Toulon (France). This cable connects through a junction box 12 lines which are separated by 60--70~m and vertically suspended by a buoy. Each line has 25 floors spaced by 14.5~m, except for line 12 which has only 20 floors. A floor consists of a triplet of optical modules (OMs) each one housing a photomultiplier (PMT) facing 45\degree\ downwards. The full detector is a tri-dimensional array of 885 PMTs \cite{antares1} \cite{antares2} which was completed in 2008 when the last line was connected.

%The neutrinos are detected via the Cherenkov light induced by the relativistic muons produced in the detector surroundings. The Cherenkov photons are detected in the array of PMTs where their arrival time and amplitude are digitized (hits) \cite{Electronics} and sent to the shore station for muon reconstruction and physics analysis. From the reconstructed muons there are two sources of background, i.\,e.\ not cosmic neutrinos: one are the atmospheric neutrinos produced in the cosmic rays (CRs) in the upper part of the Earth's atmosphere and the other are the atmospheric muons from CRs that achieve to reach the detector from the atmosphere above the detector. Only charged current interactions of neutrinos and antineutrinos were considered.

%The neutrinos are detected via the Cherenkov light induced by the relativistic muons produced in the detector surroundings. The Cherenkov photons detected by the array of PMTs are digitized (hits) \cite{Electronics} and sent to the shore station for muon reconstruction and physics analysis. From the reconstructed muons there are two sources of background: the atmospheric neutrinos produced in the cosmic rays (CRs) and the atmospheric muons from the CRs above the detector.

Neutrinos are detected via the Cherenkov light induced by relativistic muons produced in the detector surroundings by CC interactions of muon neutrinos with nuclei in water. The signals from the Cherenkov photons detected by the PMTs are digitized (`hits') \cite{Electronics} and sent to the shore station for reconstruction and physics analysis.

\section{Time-dependent analysis}
\label{analysis}
\subsection{Analyzed data}
\label{data}
The analyzed data correspond to the period from Sep\-tem\-ber 6th to December 31st, 2008 (54720--54831 modified Julian day). Some periods have been excluded from the selection in order to avoid conditions with high optical backgrounds due to the bioluminescence activity, leaving 60.8~days of life-time.

While most of the atmospheric muon background is suppressed by selecting only up-going events, a fraction of mis-reconstructed events still remains which have to be rejected in a different way. In the track-reconstruction process an algorithm based on the maximization of a likelihood is used. The likelihood function is built from the difference between the expected and the measured arrival times of the hits from the Cherenkov photons emitted along the muon track. Its maximization considers the Cherenkov photons that scatter in the water and the additional photons that are generated by secondary particles (e.\,g.\ electromagnetic showers created along the muon trajectory). A good measure of the track-fit quality is the value of the log-likelihood per degree of freedom ($\lambda$), providing a useful tool to reject the mis-reconstructed muons by applying a cut in that $\lambda$ value. This cut leaves the atmospheric neutrinos as the dominant background. An additional cut requiring the error estimated for the reconstructed muon track direction to be less than 1\degree\ is also applied in the selection of events.

\begin{figure}[!t]
  \vspace{5mm}
  \centering
  \includegraphics[width=7cm]{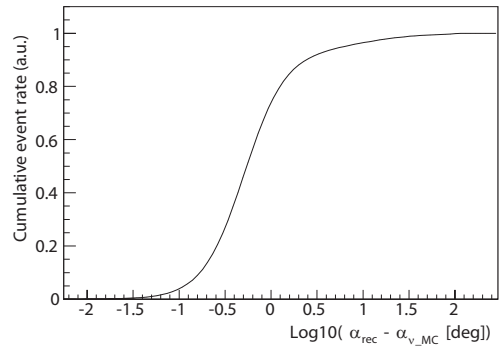}
  \caption{Cumulative distribution of the angle between the true Monte Carlo neutrino direction and the reconstructed muon direction for $E^{-2}$ upgoing neutrino events selected for this analysis. }
  \label{fig:cumulative}
 \end{figure}

The angular resolution has been estimated by Monte Carlo simulations. The cumulative distribution of the angular difference between the reconstructed muon direction and the neutrino direction, with an assumed spectrum proportional to ${E_{\nu}}^{-2}$, where $E_{\nu}$ is the neutrino energy, is shown in Fig.~\ref{fig:cumulative}. For the studied period, the median resolution is estimated to be 0.4~$\pm$~0.1 degrees.

\subsection{Time-dependent search algorithm}
\label{algorithm}
An unbinned method based on a likelihood ratio maximization has been used to perform this time-dependent point-source analysis. The data is parameterized as a two components mixture of signal and background. The goal is to determine, at a given point in the sky and at a given time, the relative contribution of each component and to calculate the probability to have a signal above a given background model. The likelihood ratio $\lambda$ is the ratio of the probability density for the hypothesis of background and signal ($H_{\mathrm{sig}+\mathrm{bkg}}$) over the probability density of only background ($H_{\mathrm{bkg}}$):
\begin{equation}
\begin{split}
\lambda &= \sum_{i=1}^{N} \log\frac{P(x_{i}|H_{\mathrm{sig}+\mathrm{bkg}})}{P(x_{i}|H_{\mathrm{bkg}})}%\nonumber
\end{split}
\end{equation}
\begin{equation}
\begin{split}
%\lambda
 &= \sum_{i=1}^{N} \log\frac{\frac{n_{\mathrm{sig}}}{N}P_{\mathrm{sig}}(\alpha_{i},\delta_{i},t_{i})+(1-\frac{n_{\mathrm{sig}}}{N})P_{\mathrm{bkg}}(\alpha_{i},\delta_{i},t_{i})}{P_{\mathrm{bkg}}(\alpha_{i},\delta_{i},t_{i})}\nonumber
\end{split}
\end{equation}
where $n_{\mathrm{sig}}$ and $N$ are respectively the unknown number of signal events and the total number of events in the considered data sample. $P_{\mathrm{sig}}(\alpha_{i},\delta_{i},t_{i})$ and $P_{\mathrm{bkg}}(\alpha_{i},\delta_{i},t_{i})$ are the probability density function (PDF) for signal and background respectively, and $\delta_{i}$ is the declination of the studied source. For a given event $i$, $t_{i}$ and $\alpha_{i}$ represent the time of the event and the angular difference between the coordinate of this event and the studied source.

\begin{figure}[!t]
  \vspace{5mm}
  \centering
  \includegraphics[width=7cm]{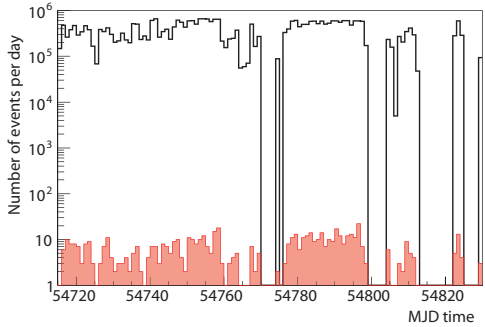}
  \caption{Time distribution of the reconstructed events. Black: distribution for all reconstructed events. Red filled: distribution of selected upgoing events ($\lambda > -5.4$ and $\beta < 1\degree$). If 0, there are no data available (i.\,e.\ detector in maintenance) or the data have a very poor quality (high bioluminescence or bad calibration). }
  \label{fig:timedist}
 \end{figure}

$P_{\mathrm{sig}}(\alpha_{i},\delta_{i},t_{i})$ and $P_{\mathrm{bkg}}(\alpha_{i},\delta_{i},t_{i})$ are described by two components:
\begin{equation}
\begin{split}
P_{\mathrm{sig}}(\alpha_{i},\delta_{i},t_{i}) = P_{\mathrm{sig}}^{\mathrm{dir}}(\alpha_{i},\delta_{i}) \times P_{\mathrm{sig}}^{\mathrm{time}}(t_{i})
\end{split}
\label{eq:psig}
\end{equation}
\begin{equation}
\begin{split}
P_{\mathrm{bkg}}(\alpha_{i},\delta_{i},t_{i}) = P_{\mathrm{bkg}}^{\mathrm{dir}}(\alpha_{i},\delta_{i}) \times P_{\mathrm{bkg}}^{\mathrm{time}}(t_{i})\nonumber
\end{split}
\end{equation}
where $P_{\mathrm{sig}}^{\mathrm{dir}}$ is the probability to have a signal event from the studied source, computed via Monte Carlo, $P_{\mathrm{sig}}^{\mathrm{time}}$ is the probability of a neutrino event derived from the gamma-ray light curve correlation, and $P_{\mathrm{bkg}}^{\mathrm{dir}}$ and $P_{\mathrm{bkg}}^{\mathrm{time}}$ are the corresponding background PDFs. The shape of the time PDF for the signal event, $P_{\mathrm{sig}}^{\mathrm{time}}$, is extracted directly from the gamma-ray light curve assuming that it is proportional to the gamma-ray flux. For the signal event directional PDF, $P_{\mathrm{sig}}^{\mathrm{dir}}$, the one-dimensional point-spread function is used, which is the probability density of reconstructing an event at an angular distance $\alpha_i$ from the true source position. Then, the directional, $P_{\mathrm{bkg}}^{\mathrm{dir}}$, and time PDF for the background, $P_{\mathrm{bkg}}^{\mathrm{time}}$, are derived from the data using respectively the observed declination distribution of selected events in the sample and the observed time distribution of all the reconstructed muons. The latter distribution (not normalized to 1) is shown in Fig.~\ref{fig:timedist}, where periods without data (due to detector maintenance, etc) or with very poor quality data (high bioluminescence activity or bad calibration) are shown as empty bins.

The $\lambda_{\mathrm{data}}$ value obtained from the data is then compared to the distribution of $\lambda$ given by the null hypothesis ($n_{\mathrm{sig}}=0$). The comparison of $\lambda_{\mathrm{data}}$ with the background only $\lambda$ distribution is used to reject the null hypothesis, where the confidence level corresponds to the fraction of the scrambled trials above $\lambda_{\mathrm{data}}$ (p-value). The discovery potential is then defined as the average number of signal events required to achieve a p-value lower than the equivalent to $5\sigma$ in 50\% of trials.

\begin{figure}[!t]
  \vspace{5mm}
  \centering
  \includegraphics[width=7cm]{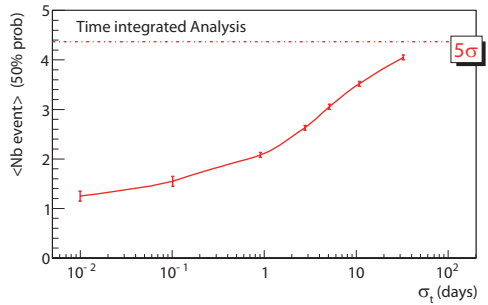}
  \caption{Average number of events required for a 5$\sigma$ discovery (50\% C.L.) of a source located at a declination of -40\degree\ as a function of the width of the flare period (solid line), a simple Heavyside function. This is compared to the number required for a time-integrated search (dashed line). The average typical length of the studied source flares goes from 1 up to 100 days. }
  \label{fig:discovery}
 \end{figure}

The average number of events required for a $5\sigma$ discovery (50\% C.L.) produced in one source located at a declination of -40\degree\ as a function of the total width of the flare period is shown in Fig.~\ref{fig:discovery}. A comparision with the numbers obtained with a standard time-integrated point source search, shows an improvement of the discovery potential by about a factor 2--3.

\subsection{Search for neutrino emission from gamma-ray flares}
\label{flare}
The sources for which this time-dependent analysis has been applied have been selected from the Fermi blazar sources reported in the first year Fermi LAT catalogue~\cite{fermicat} and in the LBAS catalogue (LAT Bright AGN sample~\cite{agn}). The selection has been done based on their variability, brightness and visibility: sources visible to ANTARES with a significant time-variability in the studied period and a flux in the high state greater than $20\times10^{-8}$~photons~cm$^{-2}$s$^{-1}$ above 300~MeV in the averaged 1 day-binned light curve were selected. The final selection list of sources includes four BLLacs and six flat spectrum radio quasars (FSRQ), listed in Table~\ref{tab:sources}. Table~\ref{tab:sources} lists their fluxes, visibility and live times.

\begin{table*}
%\centering
\begin{center}
\begin{tabular}{lccccccccc}
\hline
         Source
         & OFGL name 
         & Class 
         & Redshift 
         & $F_{300}$
         & Visibility
         & Live Time
         & $N(5\sigma)$
         & $N_{\mathrm{obs}}$
         & Fluence \\
\hline
PKS 0208-512  & J0210.8-5100  & FSRQ  & 1.003 & \ 4.43 & 1.00 & \ 8.8 & 4.5 & 0 & \ 2.8 \\
AO 0235+164   & J0238.6+1636  & BLLac & 0.940 & 13.19  & 0.51 & 24.5  & 4.3 & 0 & 18.7  \\
PKS 0454-234  & J457.1-2325   & FSRQ  & 1.003 & 13.56  & 0.63 & \ 6.0 & 3.3 & 0 & \ 2.9 \\
OJ 287        & J0855.4+2009  & BLLac & 0.306 & \ 2.48 & 0.39 & \ 4.3 & 3.9 & 0 & \ 3.4 \\
WComae        & J1221.7+28.14 & BLLac & 0.102 & \ 2.58 & 0.33 & \ 3.9 & 3.8 & 0 & \ 3.6 \\
3C 273        & J1229.1+0202  & FSRQ  & 0.158 & \ 8.68 & 0.49 & \ 2.4 & 2.5 & 0 & \ 1.1 \\
3C 279        & J1256.1-0548  & FSRQ  & 0.536 & 15.69  & 0.53 & 13.8  & 5.0 & 1 & \ 2.8 \\
PKS 1510-089  & J1512.7-0905  & FSRQ  & 0.360 & 28.67  & 0.55 & \ 4.9 & 3.8 & 0 & \ 2.8 \\
3C 454.3      & J2254.0+1609  & FSRQ  & 0.859 & 24.58  & 0.41 & 30.8  & 4.4 & 0 & 23.5  \\
PKS 2155-304  & J2158.8-3014  & BLLac & 0.116 & \ 7.89 & 0.68 & \ 3.1 & 3.7 & 0 & \ 1.6 \\
\hline
\end{tabular}
\end{center}
\caption{List of the bright, variable Fermi blazars selected for this analysis. $F_{300}$ is the gamma-ray flux above 300~MeV in $10^{-8}$~photons~cm$^{2}$s$^{-1}$. Live Time is the effective time of observation of the source in days. $N(5\sigma)$ are the number of neutrino events needed to be detected in ANTARES with $5\sigma$, while $N$ is the number of events observed in the source's direction and flare time window. Fluence is the upper limit (90\% C.\,L.) on the neutrino fluence in GeV~cm$^{-2}$, calculated according to the classical (frequentist) method for upper limits \cite{neyman}. }
\label{tab:sources}
\end{table*}
%\label{tab:visibility}
%\label{tab:results}

\begin{figure}[!t]
  \vspace{5mm}
  \centering
  \includegraphics[width=8.2cm]{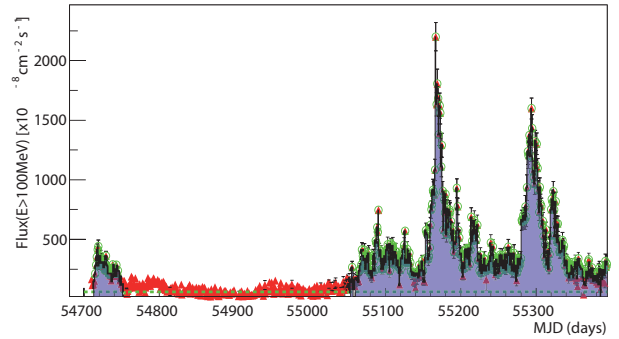}
  \caption{Gamma-ray light curve (red dots) of the blazar 3C454.3 measured by the LAT instrument on board the Fermi satellite above 100~MeV for almost two years of data. Blue histogram: high state periods. Green line and dots: baseline and significant dots above this baseline used for the determination of the flare periods. }
  \label{fig:lightcurve}
 \end{figure}

The light curves used for $P_{\mathrm{sig}}^{\mathrm{time}}$ in~(\ref{eq:psig}) are those published on the Fermi web page for monitored sources~\cite{sources} (one-day-binned time evolution of the average gamma-ray flux above a threshold of 100~MeV). Figure~\ref{fig:lightcurve} shows the high state periods (blue histogram) obtained for the Fermi LAT gamma-ray light curve of 3C454 for almost two years of data. These states are defined using a simple and robust method based on three steps:
\begin{itemize}
\item First, the baseline and its error is determined with an iterative linear and gaussian fits. After each fit, the points where the flux is above the fitted base line plus one sigma are suppressed. This is done 3 times.
\item Then all points (green dots in Fig.~\ref{fig:lightcurve}) where the measured flux minus its error is above the baseline plus two times the baseline sigma and at the same time the measured flux value is above the baseline plus three times the baseline sigma, are used as priors from which the flares are defined.
\item Then, for each selected prior, the adjacent points for which the emission is compatible with a flare are added, i.\,e.\ the adjacent points which have their flux minus their error above the baseline plus one sigma. Finally, an additional delay of 0.5 days is added before and after the flare in order to take into account the 1-day binning of the light curve. Hence, a flare has a minimum width of 2 days.
\end{itemize}
Assuming that the neutrino emission follows that in gamma-rays, the signal time PDF is simply this de-noised light curve after normalization.

\section{Results}
\label{results}
The most significant source is 3C279, which has a pre-trial p-value of 1.03\%. The unbinned method finds one high-energy neutrino event located 0.56\degree\ from the source location during a large flare in November 2008 (Fig.~\ref{fig:lightcurve}). This event has been reconstructed with 89 hits spread on 10 lines with a track fit quality $\lambda=-4.4$ and an error estimate of $\beta=0.3\degree$. The post-trial probability is computed taking into account the ten searches. The final probability of 10\% is compatible with a background fluctuation. Other source results are sumarized in Table~\ref{tab:sources}.

 \begin{figure}[!t]
  \vspace{5mm}
  \centering
  \includegraphics[width=8.1cm]{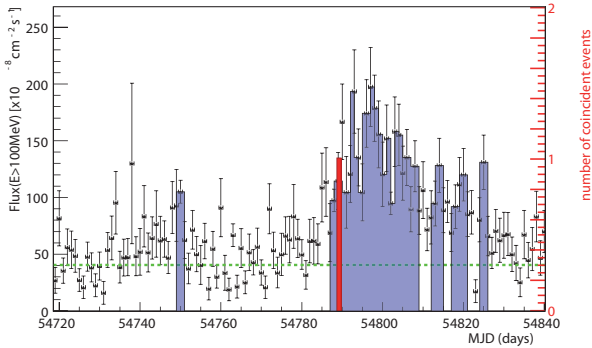}
  \caption{Gamma-ray light curve (black dots) of the blazar 3C279 measured by the LAT instrument on board the Fermi satellite above 100~MeV. Blue histogram: high state periods. Green dashed line: fit of a baseline. Red histogram: time of the ANTARES neutrino event in coincidence with 3C279. }
  \label{fig:pval}
\end{figure}

\section{Conclusion} 
\label{conclusion}
The first time-dependent search for cosmic neutrinos in ANTARES has been presented. It has used data taken with the full 12-line ANTARES detector during the last four months of 2008. Time-dependent searches are significantly more sensitive than a standard point-source search for a variable source. This search has been applied to ten bright and variable Fermi LAT blazars. The most significant observation of a flare is 3C279 with a p-value of about 10\% after trials for which one neutrino event has been detected in time and space coincident with the gamma-ray emission. Limits on the neutrino fluence have been obtained for the ten selected sources. The most recent measurements of Fermi in 2009-11 show very large flares yielding a more promising search for neutrinos \cite{flares}, and ongoing analyses are being performed together with improvements in the denoising of the light curves.

\section{Acknowledgments}
\label{acknowledgments}
%Are they necessary? Then thanks a lot to the Erlangen's police agents, for their patience with a bloody foreigner whose wallet was stolen at Zircle club the previous night, and specially for the nice girl of the boutique hotel that was my interpreter out of her work time... such a good memories. Thanks also to the bacteria or virus responsible to my 39 ºC fever during 3 days just 5 days before the deadline for the proceeding revision and for began a stay of 4 months aboard: without you my life could have been boringly easy those days. And thanks to the referee for the patience and nice corrections... I wanna holidays, wrong profession.
We gratefully acknowledge the financial support of the Spanish 
Ministerio de Ciencia e Innovaci\'on (MICINN), grants 
FPA2009-13983-C02-01, ACI2009-1020 and Consolider MultiDarkCSD2009-00064 
and of the Generalitat Valenciana, Prometeo/2009/026.

\end{document}